\documentclass[sigconf,authorversion,nonacm]{acmart}
\usepackage{multirow}

\AtBeginDocument{%
  \providecommand\BibTeX{{%
    \normalfont B\kern-0.5em{\scshape i\kern-0.25em b}\kern-0.8em\TeX}}}

\copyrightyear{2022}
\acmYear{2022}
\setcopyright{acmcopyright}\acmConference[WWW '22]{Proceedings of the ACM Web Conference 2022}{April 25--29, 2022}{Virtual Event, Lyon, France}
\acmBooktitle{Proceedings of the ACM Web Conference 2022 (WWW '22), April 25--29, 2022, Virtual Event, Lyon, France}
\acmPrice{15.00}
\acmDOI{10.1145/3485447.3512102}
\acmISBN{978-1-4503-9096-5/22/04}

\begin{document}

\title{An Empirical Investigation of Personalization Factors on TikTok}


\author{Maximilian Boeker}
\affiliation{%
 \institution{University of Zurich}
  \country{Switzerland}}
  \affiliation{%
 \institution{Technical University of Munich}
  \country{Germany}}
\email{boekermax@gmail.com}

\author{Aleksandra Urman}
\affiliation{%
 \institution{University of Zurich}
  \country{Switzerland}}
\email{urman@ifi.uzh.ch}

\renewcommand{\shortauthors}{Boeker \& Urman}

\begin{abstract}
  TikTok currently is the fastest growing social media platform with over 1 billion active monthly users of which the majority is from generation Z. Arguably, its most important success driver is its recommendation system. Despite the importance of TikTok's algorithm to the platform's success and content distribution, little work has been done on the empirical analysis of the algorithm. Our work lays the foundation to fill this research gap. Using a sock-puppet audit methodology with a custom algorithm developed by us, we tested and analysed the effect of the language and location used to access TikTok, follow- and like-feature, as well as how the recommended content changes as a user watches certain posts longer than others. We provide evidence that all the tested factors influence the content recommended to TikTok users. Further, we identified that the follow-feature has the strongest influence, followed by the like-feature and video view rate. We also discuss the implications of our findings in the context of the formation of filter bubbles on TikTok and the proliferation of problematic content.
\end{abstract}

\ccsdesc[500]{Information systems~Personalization}
\ccsdesc[500]{Information systems~Collaborative filtering}
\ccsdesc[500]{Information systems~World Wide Web}

\keywords{TikTok, algorithm audit, recommender systems, personalization, social media}


\maketitle

\section{Introduction}

In September 2016, ByteDance, a Chinese IT company, has launched a short video-sharing platform Douyin. While Douyin is only available in Mainland China, a similar application, called TikTok, was rolled out by ByteDance a year later in other countries \cite{9378032}. TikTok users can upload short videos with a variety of settings and filters, search for videos based on hashtags, content or featured background sounds, or explore the videos on their "For You" page - a feed of videos recommended to users based on their activity. As of September 2021 TikTok welcomed 1 billion active users every month and was the most downloaded application of 2020 \cite{douyinofficialreport, TikTokStatisticsIqbal, BBCTikTokMostDownloaded2020, tiktok_thanks_2021} with more than 1 billion video views recorded daily in the same year \cite{TikTokStatistiksMohsin2021,Aslam2021TikTokNumbers}. On average, people use TikTok's mobile application for 52 minutes and open it from 38 to 55 times a day \cite{Aslam2021TikTokNumbers, TikTokStatisticsIqbal}. TikTok thus has by now become a major competitor for other social media and video platforms such as Instagram and YouTube, prompting them to attempt emulating TikTok's success by implementing similar features (e.g., Instagram Reels or YouTube Shorts - short videos with recommender system-based distribution). 

TikTok is different from other major social media platforms such as Facebook or Instagram in one key aspect: its content distribution approach is purely algorithmic-driven, unlike other social media platforms where relationships between users play an important role in content distribution \cite{FacebookOfficialPostNewsFeed, alvarado2020middle, bandy2021more, klug2021trick}. Tiktok's success is largely attributed to its recommendation algorithm behind the selection of videos on the "For You" page \cite{zhao2021analysisdouyin}. The proliferation of folk theories about the innerworkings of TikTok's algorithm among its users\cite{klug2021trick}, and the appearance of several media articles and blog posts attempting to describe how the algorithm works (e.g., \cite{Scanlon2020TikTokAlgo,HowritzSeetharaman2020WSJ}) highlight public attention to TikTok's recommendation system (RS). In part, this is driven by the curiosity of users and the public and by the willingness of content creators to figure out how to achieve popularity on TikTok. Beyond that, interest in TikTok's algorithm is warranted by societal concerns such as the formation of filter bubbles and facilitation of addiction to the platform, especially among younger people as the majority of TikTok's users is between 10 and 29 years old \cite{TikTokStatisticsIqbal, basch2020covid}.

Despite TikTok's rapid growth in popularity and, consequently, its potentially high impact in political, social and cultural realms, both in part facilitated by its RS, the exact innerworkings of TikTok's RS remain a "black box" \cite{zhao2021analysisdouyin, heuer2020users}. Several studies have highlighted the importance of examining this algorithm \cite{bandy2021problematic, heuer2020users} through algorithm auditing - the investigation of functionality and impact of an algorithm \cite{mittelstadt2016automation}. While some research contributes to this goal \cite{zhao2021analysisdouyin, klug2021trick, chen2019study} and there are several media articles discussing the algorithm \cite{Scanlon2020TikTokAlgo, Wired2020Matsakis, Wang2020TikTokAlgoHooked}, many gaps remain. This is especially the case with user-centric examination of TikTok's RS - i.e., the examination of how user actions affect recommendations of the algorithm. The only analysis going in this direction has been published by the Wall Street Journal \cite{WSJ2021TikTok}, and despite yielding interesting results it was limited in scope and not strictly scientific. We aim to address the existing research gap with a user-centric audit of TikTok's algorithm. 

We make two main contributions. First, we develop and describe a methodology for conducting user-centric algorithm auditing of TikTok's RS. Second, we examine the way in which different user actions influence TikTok's recommendations within users' "For You" feeds, and discuss the implications of our findings. Of course, there is a great variety of different user actions and characteristics that can influence the highly complex RS. In our analysis we focus on a number of those we see as most explicit: user location; user language settings; liking actions; following actions; video watching actions. Our analysis is thus not exhaustive and is rather a first step towards examining TikTok's RS. Additionally, the platform periodically introduces changes to the algorithm, thus any findings we have may be only accurate for a small time window. However, our methodology can be applied at different periods in time to trace the changes in the RS, and is applicable for the examination of platforms with features similar to TikTok's "For You" feed (e.g., YouTube Shorts or Instagram Reels).

\section{Related Work}

\subsection{Auditing Recommendation Systems}
\label{section:litreview_auditRS}
Due to the widespread application of recommendation algorithms, RS can have a serious impact on how humans receive information and ultimately perceive the world \cite{bandy2021problematic, adomavicius2019hidden, sandvig2014auditing}. At the same time, "even those who train these systems cannot offer detailed or complete explanations about them or the neural networks they utilized" \cite{alvarado2020middle}. We therefore need scientific audits that shed light into the functionality of RS \cite{napoli2018social, sinha2002role}. As highlighted in a recent systematic literature review of algorithm audits \cite{bandy2021problematic}, such studies can uncover problematic behaviors of RS and personalization algorithms such as the perpetuation of various biases \cite{BaezaYatesBiasSearch2020}, construction of filter bubbles \cite{heuer2020users,ribeiro2020auditing}, personalization and randomization effects that can lead to users' unequal access to critical information \cite{makhortykh2020search,kliman2015location,HannakMeasuringPersWebSearch2013}, and price steering\cite{hannak2014measuring} \footnote{For a detailed literature review of algorithm audits see \cite{bandy2021problematic}.}. 

There are different methodological approaches to algorithm auditing. According to \cite{sandvig2014auditing}, these are: (1) code audits, (2) noninvasive user audits, (3) scraping audits, (4) sock-puppet audits, and (5) collaborative audits. Our study falls into the fourth category as we mimic user behaviour via programmatic means, thus conducting what Sandvig et al. \cite{sandvig2014auditing} refer to as a "classic" audit and following in the footsteps of other studies that examined how user characteristics and actions affect information distribution on online platforms \cite{HannakMeasuringPersWebSearch2013,Feuz2011FirstMondayPersonalWebSearching,HaimBurstFilterBubble2018}.

\subsection{TikTok-focused research}

So far research on TikTok has been conducted along two main lines: with the focus on TikTok users and their behavior, and with the focus on TikTok as a platform, including some analysis of its algorithm. The research that falls into the first category has, for example, examined the relationships between grandchildren and grandparents on TikTok in relation to COVID-19 \cite{Nouwen2021TikTokRelationshipCOVID}, analyzed political communication on TikTok \cite{bandy2020tulsaflop,TikTokPoliticalAnalysisSerrano2020} and the ways news organizations adapt their narratives to TikTok format \cite{vazquez2020let}. In the context of our study, however, the work that focuses on TikTok as a platform with an emphasis on its RS is more relevant. 

One study has examined TikTok users' assumptions about the recommendation algorithm  \cite{klug2021trick} and found "that it is quite common for TikTok users to evaluate app activity in order to estimate the behavior of the algorithm" as well as that content creators attribute the popularity (or lack of it) of their videos to TikTok's RS, and not to the video content. This study identified three main user assumptions about what influences the recommendation algorithm of TikTok on the content supply side: video engagement, posting time, and adding and piling up hashtags \cite{klug2021trick} and then, through an empirical analysis, confirmed that video engagement and posting time lead to a higher chance of the algorithm recommending a video. A few studies also described certain technical aspects of TikTok's algorithm. For instance, it has been outlined that once a new video is uploaded to TikTok, the system assigns descriptive tags to it based on computer vision analyses, mentioned hashtags, the post description, sound and embedded texts \cite{chen2019study, Wang2020TikTokAlgoHooked, Scanlon2020TikTokAlgo}. Afterwards, RS maps the tags to the user groups that match these tags, so that the recommendation algorithm can evaluate the next video to recommend from a reduced pool of videos \cite{chen2019study}. Similarly, Zhao \cite{zhao2021analysisdouyin} concluded that ByteDance systematically categorizes a large number of content to better fit the user interests. Together with this method, ByteDance utilizes user's interest, identity, and behavior characteristics to describe a user and assign categories, creators, and specific labels to them \cite{zhao2021analysisdouyin}. Further, Zhao states that TikTok solves the matching problem of an RS in two steps. Namely, through recommendation recalling which retrieves a candidate list of items that meet user preferences and recommendation ranking which ranks the candidate list based on user preferences, item characteristics, and context \cite{zhao2021analysisdouyin}. Similar to Catherine Wang's theory about the TikTok recommendation algorithm \cite{Wang2020TikTokAlgoHooked}, Zhao hypothesizes that TikTok uses the method of partitioned data buckets to launch new content \cite{zhao2021analysisdouyin}. In order to properly distribute a video, TikTok assigns newly uploaded videos to a small relatively responsive group of users (small bucket). Once the video received reasonable feedback measured by likes, views, shares, and comments surpassing a certain threshold it will be distributed to next level bucket with different users (medium bucket). This process will be repeated until a video no longer passes the threshold or lands in the "master" bucket to be distributed to the entire TikTok user community \cite{zhao2021analysisdouyin}.

In contrast to the studies above that focus on the technical aspects of TikTok's RS innerworkings or on the possible factors that can increase the likelihood that a video will be recommended to a large pool of users, we examine the way users'\footnote{By users here and below we mean TikTok content consumers, not content creators.} actions and characteristics affect the distribution of content on their "For You" feeds. Hence our analysis is centered on the content demand side rather than supply side. While the latter has been examined by the studies mentioned above, the demand side has so far been a subject of only few journalistic \cite{WSJ2021TikTok} but not scientific investigations. 

We examine a variety of user actions and characteristics that may influence the recommendation algorithm, as noted in the Introduction. Based on the background information provided by TikTok itself regarding its RS \cite{HowTikTokRecommends2020} as well as on personalization-related research in general (e.g., \cite{ricci2011introduction,HannakMeasuringPersWebSearch2013,kliman2015location}), we outline several hypotheses regarding the influence of surveyed personalization factors (user language, locations, liking action, following action, video view rate) on the users' feeds. These can be summarized as follows:
\begin{enumerate}
    \item If one user in a pair of identical users interacts with its "For You" feed in a certain way while its twin user only scrolls through its feed, the feeds of both users will diverge.
    \item Such divergence of the two users' feeds will increase overtime.
    \item Certain personalization factors have a greater impact on the recommendation system of TikTok than others.
    \item As a user interacts with specific posts in a certain way (e.g., likes them or watches them longer), that user will be served more posts that are similar to the ones it interacted with.
    \item As one of the two users interacts with its feed in a certain way, the engagement rate of the posts recommended to that user will decrease, i.e. the number of views, likes, shares, comments of recommended posts will become smaller as the user will be served more "niche" content tailored to  the user's inferred interests rather than generally popular content.
    \item Language and Location specific: Depending on the location and language a user uses to access TikTok, the user will be served different content.
\end{enumerate}

\section{Methodology}
In this section we outline the general setup of the sock-puppet auditing experiments we conducted to assess the influence of different personalization factors on TikTok that was applicable to all experimental setups, regardless of the specific factors analyzed. Distinct factor-specific characteristics of the experimental setups are mentioned in the next section separately for each personalization factor-related experimental group. Same applies to the description of the analytical strategy.

\subsection{Data Collection}
In order to empirically test the influence of different factors on the recommendation algorithm of TikTok, we needed to create a fully controlled environment so we can isolate all the external personalization factors except the one we are testing in any given experimental setup \cite{HannakMeasuringPersWebSearch2013}. Virtual agent-based auditing (or "sock-puppet" auditing \cite{sandvig2014auditing}) is an appropriate methodology for creating such an environment while mimicking realistic user behaviour to assess the effects of different personalization factors \cite{HaimBurstFilterBubble2018,urman2021matter}. Thus, we created a custom web-based bot (virtual agent with scripted actions) that is able to log in to TikTok, scroll through the posts of its "For You" feed and interact with them, e.g. like a post. Similar to Hussein and Juneja \cite{HusseinMeasuringMisinformationVideoSearch2020}, our program ran the ChromeDriver in incognito mode to establish a clean environment by removing any noise resulting from tracked cookies or browsing history that may originate from the machine on which the bot program was executed. The source code can be accessed on GitHub \footnote{\href{https://github.com/mboeke/TikTok-Personalization-Investigation}{https://github.com/mboeke/TikTok-Personalization-Investigation}}.

The scripted actions of the bot were executed as follows: first the program initialized a Selenium Chrome Driver session\footnote{In order to obscure the automated interaction of our bot program we followed the suggestions of Louis Klimek's article \cite{Klimek2021UndetectingBot}.} with browser language set to English per default (depending on the test scenario, we adjusted the language; see details in Table \ref{tab:testgroupdetails}), navigated to the TikTok website (https://www.tiktok.com), logged in as a specific user (login verification step was completed manually; we describe how user accounts were created below), and handled a set of banners to assure an error-free interaction with the user's "For You" feed; then it scrolled through a pre-specified number of posts and executed actions such as following or liking (as scripted for a specific experiment and "run" (execution round) of the program); while scrolling through the "For You" feed, the bot retrieved the posts' metadata from the website's source code and extracted more data from the request responses. In the testing rounds ahead of the deployment of the bots we established that every time TikTok's website was accessed it automatically preloaded about 30 posts to be displayed on the "For You" feed. Hereafter we refer to such groups of 30 posts as \textit{batches}. As soon as the pre-specified number of batches\footnote{3 by default for all experiments, though for some 5 batches were collected, as noted below and in Table \ref{tab:testgroupdetails}.} was scrolled through, the bot paused the last video and terminated the ChromeDriver session once all requested data was temporally stored to avoid unintentional interaction with the TikTok's feed. Afterwards all the data was stored in a PostgreSQL database hosted on Heroku. During our experiment we operated five local machines, four ran Windows 10 Pro and one macOS; as two users that were compared with each other (see below) always ran from the same local machine, the between-machine differences had no potential effect on our results. All machines were connected to the remote database. 

For each run of the bot, we scripted a set of specifications which defined the characteristics of each run, e.g. web-browser language, test user, number of batches to scroll through etc. According to Yi, Raghavan, and Leggetter \cite{YiDiscoveringUsersGeo2009}, web services can identify a user’s location through their IP address. We therefore have assigned a dedicated proxy with a specific IP address to every test user due to three reasons: (1) every test shall be performed at a certain location, (2) to obscure the automated interaction, and (3) to link a specific IP address to a specific test user. We utilized proxies from WebShare \footnote{\href{www.webshare.io}{www.webshare.io}} and acquired phone numbers from Twilio\footnote{\href{www.twilio.com}{www.twilio.com}} to setup user accounts. We utilized user phone numbers instead of email-addresses as those would require a completion step on the mobile application. Similarly to \cite{HusseinMeasuringMisinformationVideoSearch2020, HannakMeasuringPersWebSearch2013, HannakPriceDiscrimination2014}, every test user was manually created using its dedicated proxy and incognito mode to reduce the influence of any external factors. Every machine executed one program run at a time which consisted of two bot programs being executed in parallel.

As noted in the Introduction, we aimed to establish the influence of several user actions and characteristics on TikTok's RS and thus the personalization on the platform's "For You" feed. We focus on the influence of the most explicit actions and characteristics (tested factors): following a content creator, liking a post, watching a post longer, and the language and location settings. To assess their influence on TikTok's RS, we conducted several experiments using the bot program as outlined above. We describe the experiments related to each of the tested factors below.

\subsection{Experiment Overview}
We created one experimental group with different experimental scenarios for every tested factor. For every scenario we have performed about 20 different runs which mainly consisted of two users (bots) executing scripted actions on one local machine in parallel. One of the two was the active and the other the control user. The active user performed a certain action, e.g. liking a post, while the control user only scrolled through the same number of batches as its twin user, looking at each post the same amount of seconds. We thus followed an approach similar to Hannak et al. \cite{HannakMeasuringPersWebSearch2013} and Feuz, Fuller, and Stalder \cite{Feuz2011FirstMondayPersonalWebSearching} by creating a second (control) user, that is identical to the active user except one specific characteristic/action - one of the tested personalization factors, - in order to measure the difference of the users’ feeds by comparing the meta-data of the posts that both saw. If the posts on the feeds vary and do so more than we would expect due to inherent random noise (see \cite{HannakMeasuringPersWebSearch2013}), the difference can be attributed to the personalization of the recommendation algorithm of TikTok triggered by the tested factor. Every test scenario was executed twice a day, although the execution order varied, until all 20 test runs were completed.

\subsection{Data Analysis}
In order to analyse the results of our experiment we used four different analysis approaches.

\textit{ First,} we analyzed the difference between the feeds of two users by utilizing the Jaccard Index to measure the overlaps between posts, hashtags, content creators, and sounds between that each of the users encountered on their feed. Similar to previous work on measuring personalization online \cite{HannakMeasuringPersWebSearch2013, urman2021matter}, this approach allows us to identify to which degree the user feeds differ with respect to different metrics and attribute their variation to the influential factor being tested. Additionally, we compute the change trend in the discrepancies by fitting the obtained data to a linear polynomial regression.

\textit{Second,} we analyze the number of likes, views, comments, and shares of a post. As noted by \cite{klug2021trick}, one can evaluate a post's popularity on TikTok based on these metrics. We therefore examine these attributes to evaluate the popularity of individual TikTok posts recommended to the bot users, and also trace how average popularity of posts recommended to a user changes overtime (i.e., we expect that with time due to personalization the posts recommended to a user should become more tailored to their interests thus more "niche" and less popular on the platform as a whole). 

\textit{Third,} TikTok itself \cite{TikTokOfficalRecommend2020} as well as \citep{zhao2021analysisdouyin, DominguesForensicTikTok2020} mention the importance of hashtags to the platform implying that content classification and distribution is heavily based on hashtags. We analyzed the reappearance hashtags as well as sounds and content creators on a given user's "For You" feed overtime to investigate whether TikTok picked up that user's interests as proxied by these post properties. Additionally, we cleaned the data before the analysis by removing overly common hashtags, e.g. "\#fyp" (shortcut of the "For You" page) as those mentioned too frequently would obscure the real similarity - or absence of it - between different posts. 

\textit{Fourth,} we analyzed the similarity of two posts by analyzing the semantics of those posts' hashtags using a Skip-Gram model \citep{mikolov2013efficient}.

\subsection{Ethical considerations}
TikTok's Terms of Service (ToS) explicitly prohibit content scraping for \textit{commercial purposes} \cite{tiktok_tos}. As our audit is done for academic purposes only, without any commercial applications, we do not violate TikTok's ToS. Our bots have interacted with the platform as well as with the content creators (e.g., by liking/following them). However, as we used only few agents, we did not cause any disruption to the service and had only marginal, non-intrusive and completely harmless interactions with the content creators. Our research qualified as exempt from the ethical review of the University of Zurich's OEC Human Subjects Committee according to the official checklist.

\section{Experiments}
All experiments were conducted between late June 2021 and mid-August 2021. In total, there were 39 successfully completed\footnote{Beyond those 39 there were several runs we excluded from the analysis due to technical issues-related errors in the execution that could affect the results (e.g.,  when a bot got "stuck" on one post "watching" it for a long time which could affect the behaviour of the RS in undesirable ways). Such failed runs are listed together with successful runs in the overview Table \ref{tab:testgroupdetails} for reference but their IDs are marked in red.} experimental scenarios during which we collected the data on 30’436 different posts, 34’905 distinct hashtags, 21’278 different content creators, and 20’302 distinct sounds.
In the sections to come we elaborate on the most significant findings for brevity reasons. We list all relevant details including the ID of each experimental scenario and corresponding bot users IDs in Supplementary Material in Table \ref{tab:testgroupdetails}.

\subsection{Controlling Against Noise}
As introduced in section \ref{section:litreview_auditRS}, when auditing algorithms one needs to identify potential sources of noise to assure any differences observed between users in experimental scenarios are due to personalization, and not inherent "noise" or randomization. In this section, we elaborate on the potential sources of noise and how we addressed them.

Accessing TikTok from different locations may result in different content being recommended. We control for this personalization by assigning dedicated IP addresses located within the same country and obtained from the same proxy provider for every pair of test users. As the device settings can be another influence to TikTok's RS, every machine uses the same ChromeDriver version and a proxy dedicated to a specific user to access TikTok. 

TikTok points out that their "[...] recommendation system works to intersperse diverse types of content along with those you already know you love". They specifically state that they will "interrupt repetitive patterns" to address the problem of the filter bubble \cite{TikTokOfficalRecommend2020}. We need to control for this type of noise - the difference between two feeds that is triggered by the aforementioned design choices and inherent randomization and not the tested factor. In order to account for it and other potential sources of noise in the analysis, we created 11 experimental control scenarios, where none of the two users interacts with its feed in any way in order to measure the "default" levels of two users' "For You" feed divergence. To increase the robustness of our observations, we slightly varied the conditions of the control scenarios: some of our test scenarios collected five instead of three batches, or collected data from the first few posts of a feed while others did not. Our results reveal that there is no clear correlation between the level of users' feed divergence and collecting and not collecting the first few posts or collecting three vs five batches of posts. Thus, we treat these different settings as equivalent. Nonetheless, when accounting for noise in the analysis of experimental results for different tested factors (see below), we compared the observations for each tested factor scenario only with the observations of a control scenario fully corresponding to it (e.g., in terms of the number of batches of data collected). Using the data collected from the control scenarios, we computed a "noise value" (the level of divergence of two users' feeds when the users are identical and do not interact with their feeds in any specific way) for the number of different posts, hashtags, content creators, and sounds by averaging over differences across all test runs and scenarios. The percentage of different posts, content creators, hashtags, and sounds was 66.17\%, 66.05\%, 58.62\%, and 64.47\% for all scenarios collecting five batches. For scenarios that collected three batches these percentages corresponded to 69.74\%, 68.15\%, 59.63\%, and 68.05\%.

\begin{figure*}[!tbp]
    \centering
    \begin{minipage}{0.45\textwidth}
        \includegraphics[width=\textwidth]{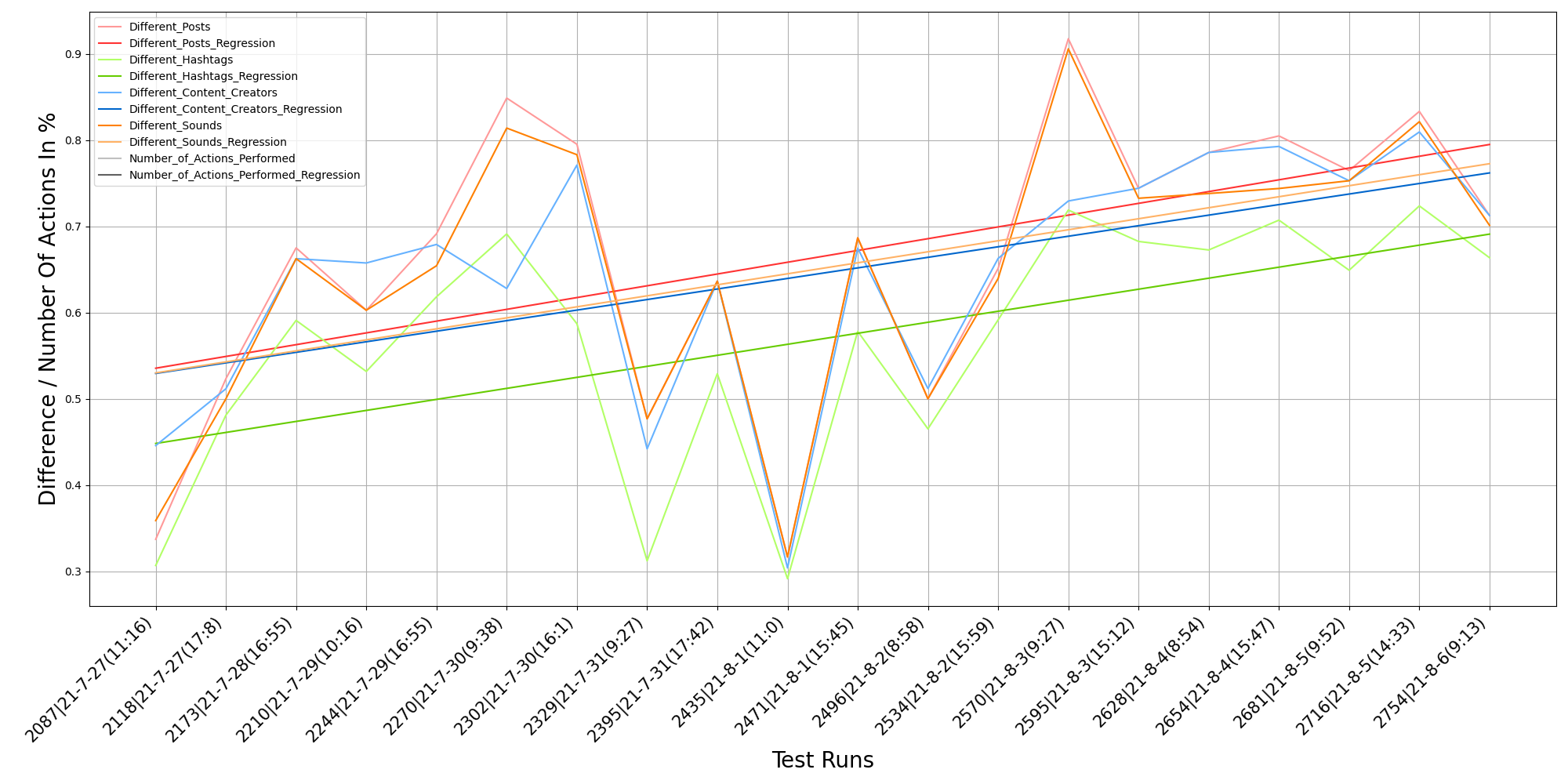}
        \caption{Difference of feeds per test run for test scenario 7 before accounting for drops.}
        \label{fig:resultsscenario7}
    \end{minipage}
    \begin{minipage}{0.45\textwidth}
        \includegraphics[width=\textwidth]{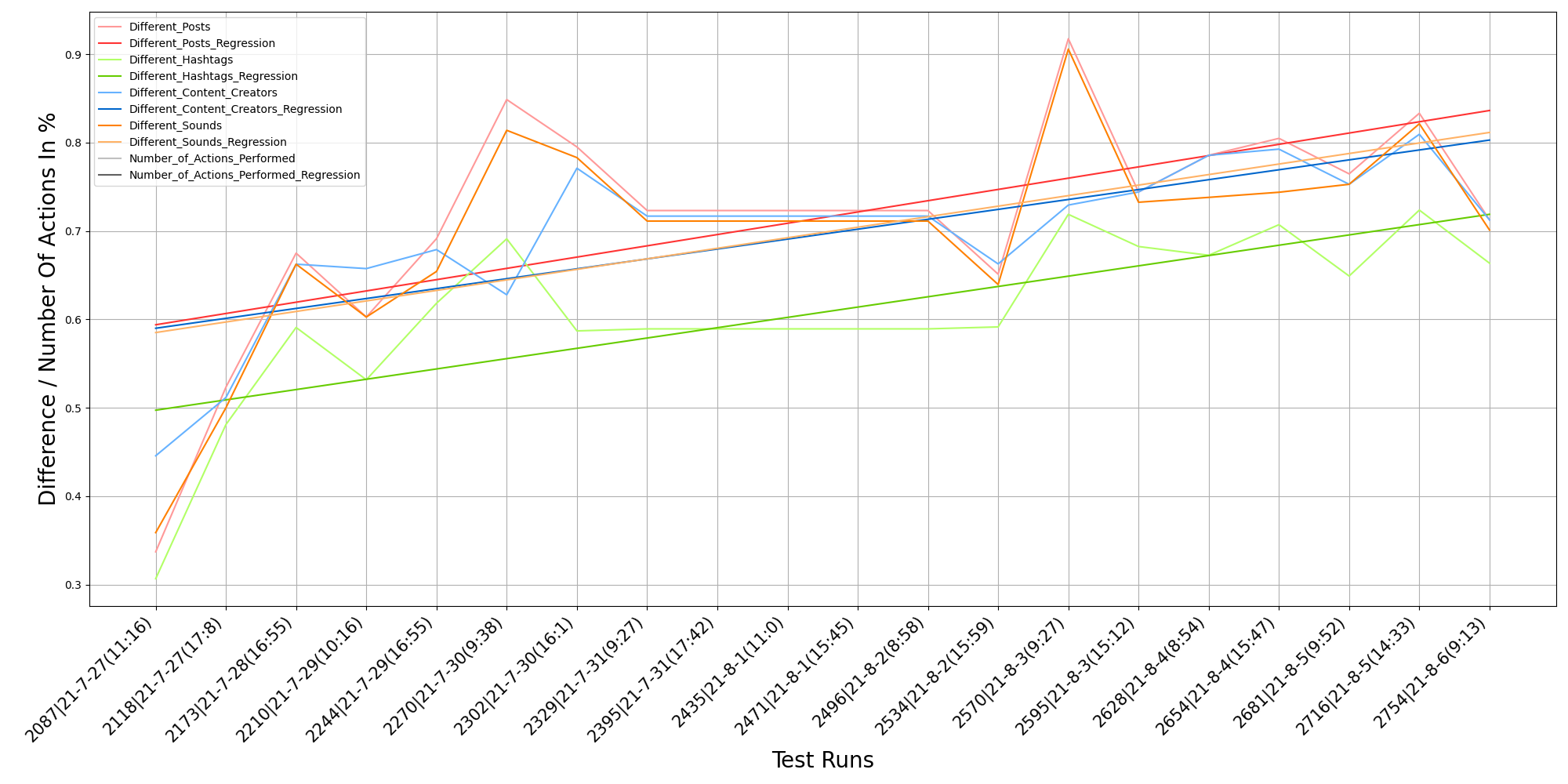}
        \caption{Difference of feeds per test run for test scenario 7 after accounting for drops.}
        \label{fig:resultsscenario7accountingdrop}
    \end{minipage}
    \label{fig:diffanalysiscontrolscenario7}
\end{figure*}

For brevity reasons here we present detailed results from only one of the 11 control scenarios (scenario ID 7), it however is similar to other control scenarios. Figure \ref{fig:resultsscenario7} shows strong fluctuations of the difference between the users' feeds, the most dominant being between test runs ID 2302 and 2534. We identified such drops in all test scenarios and figured that they regularly occur around the end of a week or weekend. Since TikTok continuously improves their recommendation algorithm \cite{TikTokOfficalRecommend2020}, we believe that these drops must be related to software releases. We therefore accounted for these (presumed) software updates by averaging the values right before and after the drops to lift the graph as shown in figure \ref{fig:resultsscenario7accountingdrop}. In figure \ref{fig:postmetricscontrolscenario7} we observe that there are huge fluctuations in the levels of popularity (as proxied by likes and views) and engagement (proxied by shares and comments) of posts recommended by the RS. TikTok's algorithm seems to prioritize popular posts in the beginning, which is likely done to provoke a user feedback and thus overcome the cold-start problem. We averaged over the slopes of the trend lines of every difference analysis approach in order to compare the control and test scenarios. The corresponding values are provided in the Supplementary Material \ref{section:differenceanalysisresults}. Hypothetically, if a tested factor indeed influences the recommendation algorithm, then the resulting feed should show stronger differences in its content than the ones of our control scenarios.

\subsection{Language and Location}
\label{section:languagelocation}
\textit{Setup.} In order to show the influence of a language of the TikTok website and location from which the user accesses the service we created four different experimental scenarios (see Table \ref{tab:testgroupdetails} for the specifications). For each of those the bot only collected data, no test user performed any action on its feed. However, bot users in each pair were either running from different locations (manipulated via proxies) or had different language settings (set up via their TikTok profiles). Comparing the number of overlapping posts between user pairs that belonged to the same scenario we were able to identify the impact of a language and location. Scenario 12 and 13 contained two test user pairs each, one accessing TikTok from the US and the other from Canada, both in English. Unfortunately, however scenario 13 was excluded due to faulty bot behavior as noted in Table \ref{tab:testgroupdetails}. Scenario 14 again consisted of two user pairs, one located in the US using English, the other in Germany with language set to German. For one user of each pair we switched the locations to Germany and the US back and forth to test if the RS "reacts" to the changes in the location immediately. In scenario 15 we focused on the influence of the language settings only. The experiment included four test user pairs. All accessed TikTok from the US, but each pair with one of the four languages: English, German, Spanish, and French. We decided to execute this experiment in the US as its population is reasonably large and according to Ryan \cite{ryan2013language} apart from English, Spanish, German, French belong to the four major languages spoken in that country. 

\begin{figure*}[!tbp]
    \centering
    \begin{minipage}{0.3\textwidth}
        \includegraphics[width=\textwidth]{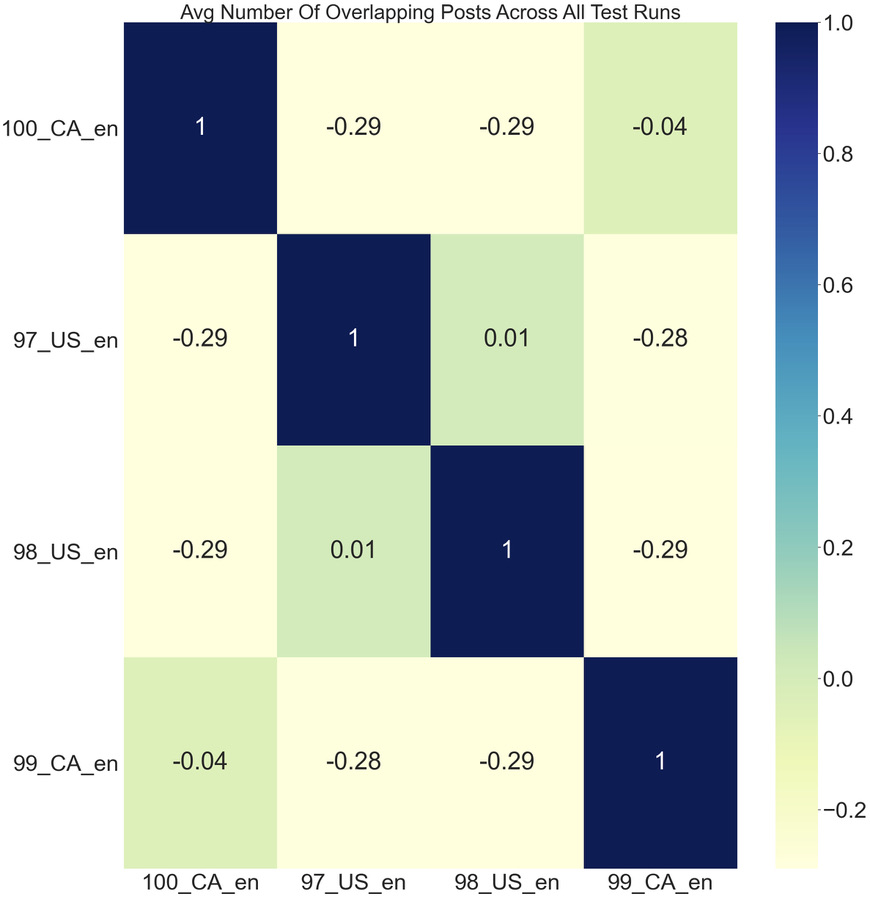}
        \caption{Results of test scenario 12.}
        \label{fig:resultsscenario12}
    \end{minipage}
    \begin{minipage}{0.3\textwidth}
        \includegraphics[width=\textwidth]{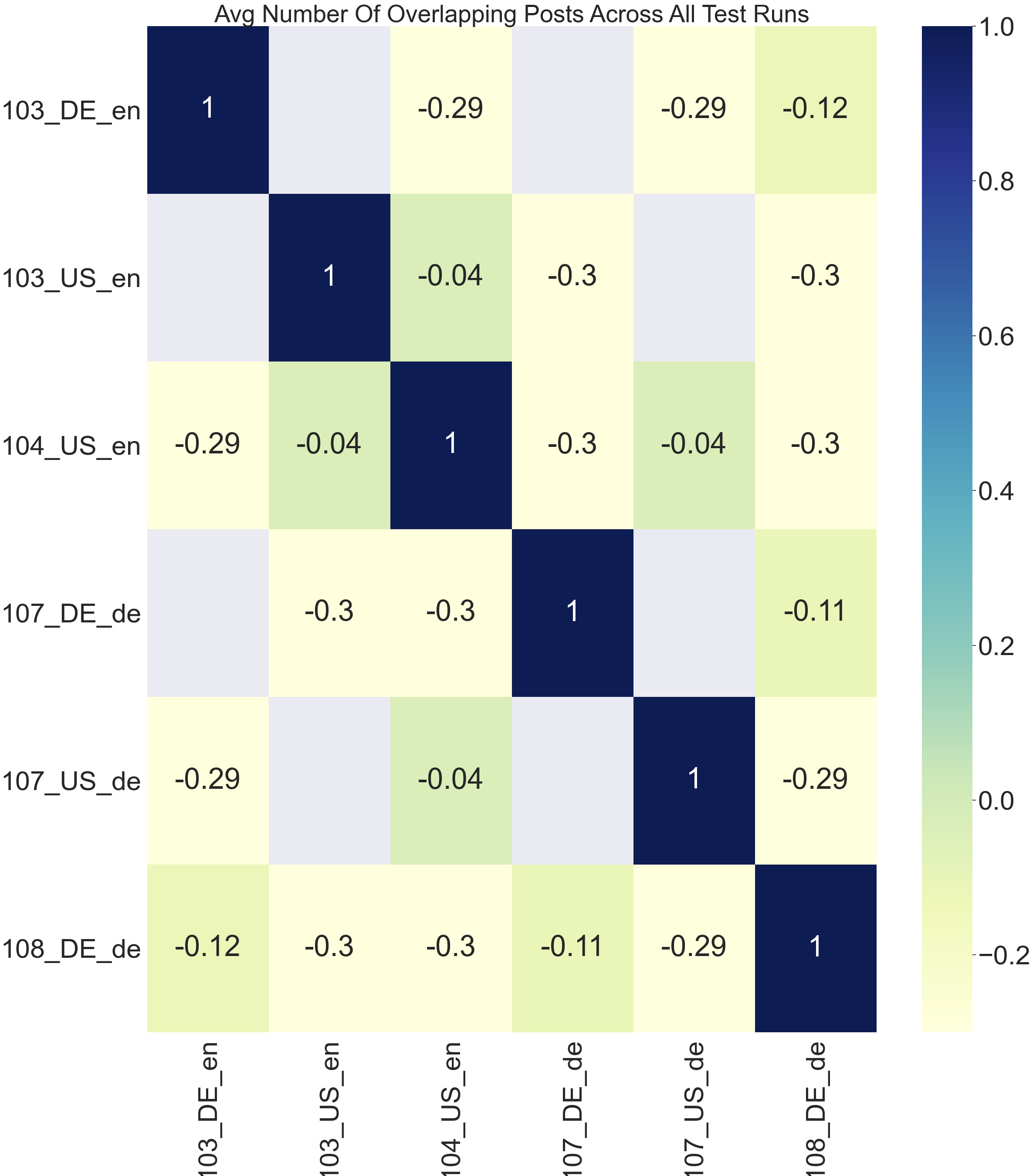}
        \caption{Results of test scenario 14.}
        \label{fig:resultsscenario14}
    \end{minipage}
    \begin{minipage}{0.3\textwidth}
        \includegraphics[width=\textwidth]{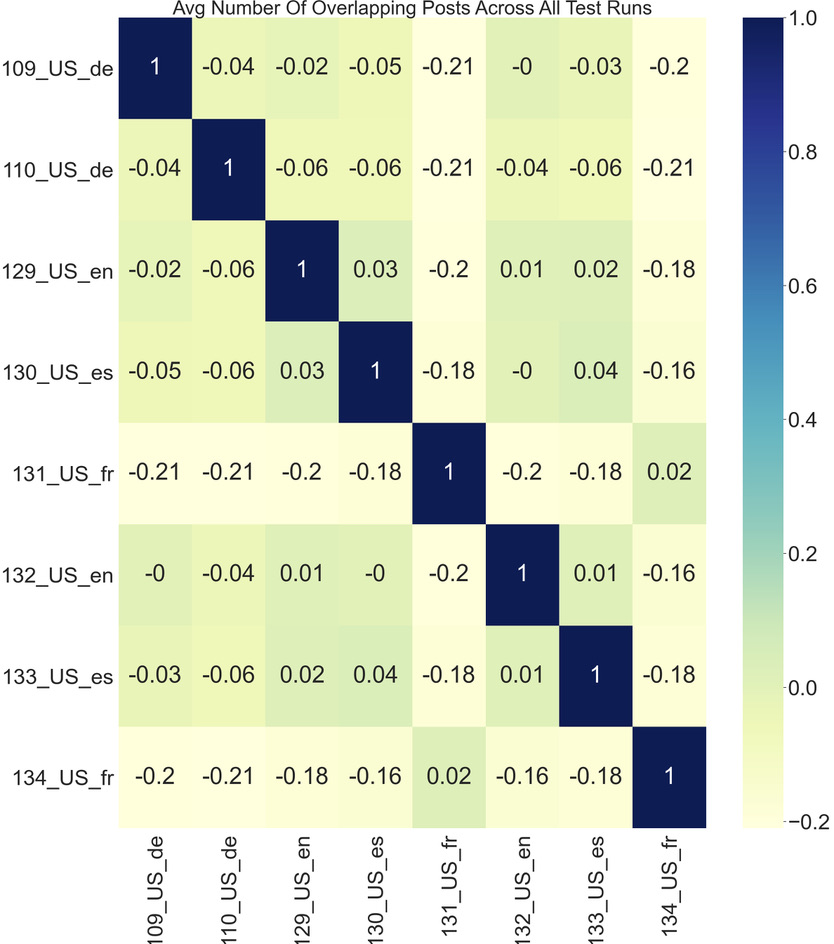}
        \caption{Results of test scenario 15.}
        \label{fig:resultsscenario15}
    \end{minipage}
    \label{fig:resultslanguagelocation}
\end{figure*}

\textit{Results.} The heat maps in Figures \ref{fig:resultsscenario12}, \ref{fig:resultsscenario14}, and \ref{fig:resultsscenario15} visualize the averaged overlapping posts of each user of each corresponding test scenario across all test runs. Note that the negative values result from accounting for the overlapping noise of 35.38\%. All three charts \ref{fig:resultsscenario12}, \ref{fig:resultsscenario14}, and \ref{fig:resultsscenario15} show that different locations have a strong impact on the posts shown by TikTok. For example, on the heat map in Fig. \ref{fig:resultsscenario12} both users 97\textunderscore{US}\textunderscore{en} and 98\textunderscore{US}\textunderscore{en} have a higher average of overlapping posts than the users 97\textunderscore{US}\textunderscore{en} and 99\textunderscore{CA}\textunderscore{en}. Figure \ref{fig:resultsscenario14} shows the same phenomenon even though the users switch their location in the meantime. This also implies that language does not influence the RS as strong as the location does. The heat map in Fig. \ref{fig:resultsscenario15} indicates that accessing TikTok using the same language setting does not always result in the highest overlap (e.g. comparing all users with 109\textunderscore{US}\textunderscore{de}). We learn that a user accessing TikTok from the US is likely to see more content in English than any other language regardless of the language settings, which makes sense as English is the country's official and most dominant language. This is the case for all examined languages except French - the feeds of users with French set as default language are more similar to each other than to users with other language settings. It seems as if TikTok interprets French to be more different to English, Spanish, and German than those three languages to each other.

\subsection{Like-Feature}
\textit{Setup.} As one of TikTok's influential factors, the like-feature could be interpreted as a proxy to understand user preferences, similar to a user rating \cite{TikTokOfficalRecommend2020, zhou2012state}. We created 11 different test scenarios incorporating different approaches of selecting the posts to like: randomly, based on user personas defined by set of hashtags\footnote{For example, the set of hashtags of user 145 of scenario 39 is the following: ["football", "food", "euro2020", "movie", "foodtiktok", "gaming", "film", "tiktokfood", "gta5", "gta", "minecraft", "marvel", "cat", "dog", "pet", "dogsoftiktok", "catsoftiktok", "cute", "puppy", "dogs", "cats", "animals", "petsoftiktok", "kitten"]. All of these hashtags correspond to very popular interests, same was true for all persona scenarios.}, and those that matched specific content creators or sounds. With regards to the persona-based selection, we followed the approach of  \cite{Feuz2011FirstMondayPersonalWebSearching} to artificially create user interests based on a set of values, in our case using hashtags as a proxy to determine whether a video matches these pre-specified interests of a user or not. If at least one hashtag of the currently displayed post would matched the pre-defined set of hashtags corresponding to user interests, the user would like the post. The above referenced Table \ref{tab:testgroupdetails} specifies which scenario followed what kind of post-picking-approach.

\textit{Results.} Overall, our analysis reveals that differences of feeds for scenarios that collected only three batches increase stronger than for the control scenarios. This, however, does not occur for scenarios that collected five batches, potentially indicating that the RS adapts the feed of a user trying to "infer" their interests even in the absence of any user actions, and this effect gets stronger the longer a user remains idle. Still, overall across all like scenarios (regardless of how the liking actions were specified), the users' feeds diverged stronger than in the control scenarios (as depicted in Table \ref{tab:likeaveragedresults}). That being said, the feeds in the scenarios for which active users were defined by only very few common hashtags did not diverge very much. We propose to run additional tests in future work with more specific, niche hashtags to investigate their feed change. Again we focus on scenario 21 as an example and omit details of the remaining scenarios for brevity reasons. The analysis of the feed difference and post metrics for scenario 21 reveal that the feeds become more different, show less popular posts in terms of likes and vies, and thus, imply that more personalized posts are fed to the active users than its twin control user. Similarly, the hashtag similarity analysis of scenario 21 reveals that the feed of user 123 becomes similar faster than that of control user 124. Also, the test scenarios where active users liked only certain content creators (scenarios 23 \& 24) or sounds (25 \& 26) showed a higher increase in differences compared to the appropriate control scenarios. The analysis of reappearing content creators or sounds for these scenarios also show that the content creators or sounds for which a post was liked reappeared more often than others.

We conclude that liking posts does influence the recommendation algorithm of TikTok. However, we figured that an arbitrary selection of posts to like does not have as strong an effect as persona-based picking, or based on a specific set of content creators or sounds.

\subsection{Follow-Feature}
\textit{Setup.} We created six different test scenarios to test the follow-feature. For each one of them one of the user pairs followed only one random content creator every other test run. Again we had to exclude the scenario 29 as the bot got stuck.

\begin{figure}[h]
    \centering
    \includegraphics[width=0.3\textwidth]{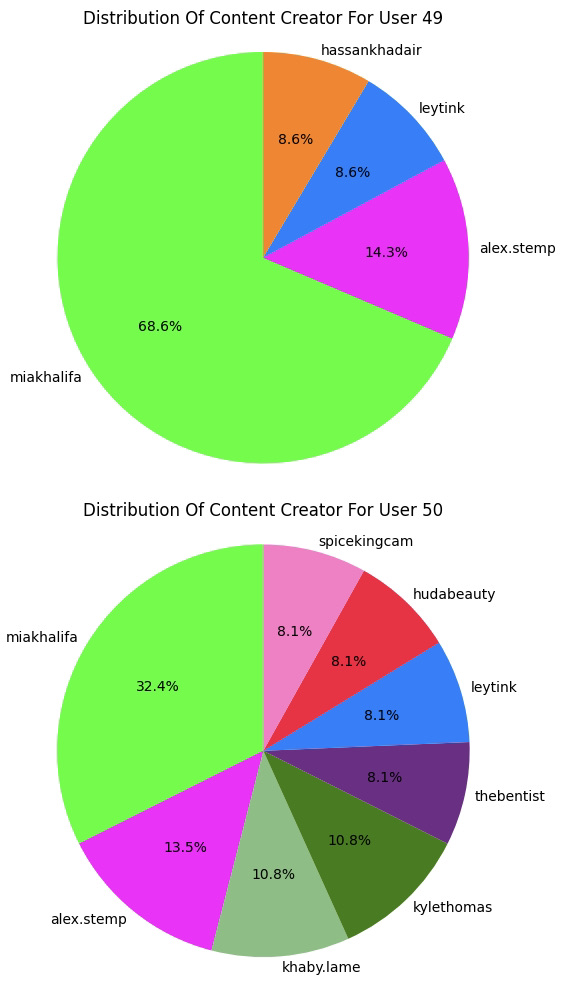}
    \caption{Distribution of content creators across all test runs for scenario 28.}
    \label{fig:contentcreatorappearancescenario28}
\end{figure}

\textit{Results.} Our overall difference analysis as well as the hashtag similarity analysis let us conclude that following a certain content creator undoubtedly influences the recommendation algorithm (details in Table \ref{tab:followaveragedresults}). Figure \ref{fig:contentcreatorappearancescenario28} related to scenario 28 further underpins this finding by displaying a greater variance of content creators for the control user 50 than the active user 49. Interestingly, three out of four content creators most frequently encountered by user 49 are not followed by this user. We suggest this might be due to their similarity to the creators followed by user 49 coupled by overall popularity (but not the latter alone as otherwise we would expect them to pop up in the control user's feed with similar frequency). However, our hashtag similarity analysis of scenario 28 shown in figure \ref{fig:followresultsdiffhashtagsim} again illustrates a strong influence of the follow-feature as the posts of the active user's feed become similar to each other faster than those in the feed of the control user (21\% > 18\%).

\subsection{Video View Rate}
\textit{Setup.} With YouTube's design change in its recommendation algorithm that introduced accounting for the percentage a user watched a video, the overall watch time on the platform started rising by 50\% a year for the next three years \cite{NewtonYouTubeFeed2017}. Google calls this metric the "video viewership" which measures the percentage that was watched of a certain video \cite{YouTubeHelpArticle}. Given the importance of the feature on YouTube, we hypothesized it might also be relevant for the TikTok's RS system and set out to test this. We adjusted the "video viewership" metric as describe by Google to our purposes and call it the video view rate (VVR). We created ten different experimental scenarios to examine the influence of the VVR on TikTok's recommender system. The set of experimental scenarios was equally split into five that randomly picked posts and the other five based on a user persona. For both groups of test scenarios the share of video length that the bot users "watched" was varied between 25\% and 400\% (400\% = watching a video four times), the details for each scenario are listed in Supplementary Material Table \ref{tab:testgroupdetails}.

\textit{Results.} Our analysis depicted in Table \ref{tab:vvraveragedresults} reveals that the feed difference of the persona scenarios (those that "selected" videos to watch longer based on pre-specified sets of hashtags) increases significantly stronger than for other VVR scenarios allowing us to conclude that the TikTok recommendation algorithm reacts stronger to the VVR differences based on specific user profiles (the more niche the better) than on user profiles that randomly pick posts. Our results from the like-feature test scenarios align with these findings. Contrary to our assumptions, the feeds of scenario 33 with the active user watching only 25\% of certain posts increase stronger in their difference than for scenario 35 with the active user watching 75\% (averaged difference 0.85\% > 0.56\%). We observe the same with scenario 38 (active user watching 50\%) and 40 (active user watching 100\%). One explanation might be that TikTok RS "assumes" users decide within the first 25\% (or 50\% respectively) of the video duration whether they like the video or not. The remaining time is thus no longer relevant. Another reason may be that the feeds of scenario 33 just happened to be slightly more different from the beginning, and therefore, changed faster. Or the feed of user 77 may be more volatile than of user 81 as user 77 watches only 25\% resulting in TikTok serving many different videos. Yet another explanation may be that watching 75\% instead of 25\% sends a stronger negative feedback. Looking at the hashtag semantics of the feeds for both scenarios reveals that the similarity of the feed from user 81 (slope: 10.92\%) increases a lot faster than for user 77 (slope: 7.79\%). Likewise, the hashtag similarity for user 91 (slope: 16.03\%) grows quicker than for user 87 (slope: 7.98\%). An additional indicator of personalization within the VVR tests that involve user personas is the number of posts that were watched longer as well as the time a bot needed to complete a test run. Our analysis revealed that user 91 watches increasingly more posts for an extended time frame with an average duration of 33.73 minutes than user 87 with an average duration of only 27.78 minutes.

Even though the feed difference analysis appears to increase stronger for users who watch less of a post, our findings allow us to conclude that not only watching a video longer than others influences the recommendations of TikTok's algorithm, but also the longer one watches the stronger it influences the algorithm.

\subsection{Concluding Results}
In this section we summarize the findings with respect to the previously introduced hypotheses. For the majority of all experimental non-control scenarios, the feeds become more different and continue to do so as the active user continues interacting with its feed (hypothesis 1 and 2). Furthermore, our data reveals that certain factors influence the recommendation algorithm of TikTok stronger than others. The order of the most influential factor to the least among those that were tested is the following: (1) following specific content creators, (2) watching certain videos for a longer period of time, and finally (3) liking specific posts. Interestingly, the influence of the video view rate is only marginally higher than the one of the like-feature. The number of performed and fully completed test scenarios as well as the number of collected batches may be one of the reasons. Another one may be the approaches to picking a post to interact with: on the one hand random picking of posts, which was identified as not a strong influential factor, and on the other persona-based picking, where the user were defined by very common and similar hashtags. The fact that watching a post for a longer period of time has a greater effect on TikTok’s recommendation algorithm than liking it aligns with TikTok's blog post \cite{TikTokOfficalRecommend2020}. However, we can not confirm the findings of the WSJ investigation \cite{WSJ2021TikTok} as our data shows that following specific content creators influences the "For You" feed stronger than all the other tested factors. Elaborating on hypothesis four (increased within-feed similarity of content served to an active user) is not as straightforward. Overall, the follow feature scenarios indicate that the RS of TikTok indeed serves to the active user more posts of the content creators the user followed. The same is true for like feature where the user liked posts of certain content creators and/or with certain sounds. However, we do not identify a clear pattern for post attributes reappearing more often than others for the like- and VVR- tests where users picked posts randomly or based on predefined sets of hashtags. The first observation may again be due to the arbitrary selection. The second might be because of the hashtags that defined the personas are very popular and, thus, appear equally often for the active and corresponding control user. We plan on addressing this issue in future work by running tests with personas being defined by more specific, niche hashtags. However, the similarity analysis of the feeds reveals that in most cases the posts in the feeds of active users became similar faster than in the feeds of control users. We therefore consider hypothesis four to be true as well. Considering the averaged slopes of the combined post metrics, the feeds of active users do not always decrease faster than for the control user. We therefore reject hypothesis 5. Even though TikTok serves more personalized content it still recommends posts with very high numbers of views, likes, shares, and comments. Section \ref{section:languagelocation} revealed that both language and location effect the TikTok posts recommended to a user (hypothesis 6).

\section{Discussion}
In the past decade algorithmic personalization has become ubiquitous on social media platforms, heavily affecting the distribution of information there. The recommendation algorithm behind TikTok's "For You" page is arguably one of the major factors behind the platform's success \cite{zhao2021analysisdouyin}. Given the popularity of the platform \cite{TikTokStatistiksMohsin2021,Aslam2021TikTokNumbers}, the fact that its largely used by younger users who might be more vulnerable in the face of problematic content \cite{weimann2020research}, as well as the central role TikTok's RS plays in the content distribution, it is important to assess how user behaviour affects one's "For You" page. We took the first step in this direction. In this section we outline the implications of our findings as well as the directions for future work.

Our analysis revealed that following action has the largest influence on the content served to the users among the examined factors. This is important since following is a conscious action, as contrasted for example to mere video viewing which could happen by accident or be affected by unconscious predispositions. One can watch something without necessarily liking what they see, especially in the case of disturbing or problematic content. Hence, according to our results users have some control over their feed through explicit actions. At the same time, we find that video view rate has a similar level of importance to the RS as liking action. This can be problematic: while likes can be easily undone and users unfollowed, one can not "unwatch" a video, thus the influence of VVR on the algorithm severely limits the users' control over their data and the behaviour of the algorithm. Given the proliferation of extremist content on the platform and TikTok's insofar insufficient measures to limit the spread of problematic content \cite{weimann2020research} as well as the high degree of randomization in the videos served to a user as identified by us, one can be potentially driven into filter bubbles filled with harmful and radicalizing content by simply lingering over problematic videos for a little bit too long. To alleviate this, we, similarly to \cite{weimann2020research,zhao2021analysisdouyin}, suggest that TikTok should do more to filter out problematic content. Additionally, the platform could provide users with more options to control what appears in their feeds. For example, TikTok could add a list of inferred user interests available for control and adjustments to the user itself. TikTok already enables its users to update their video interests via settings, but only within few superficial categories. We suggest to provide a consistently updated list of inferred user interests using very detailed content categories based on which the user can always identify which interests the TikTok RS inferred from their interaction with the app. The user should also be able to adjust the list. According to \citep{mittelstadt2016automation} and \citep{sinha2002role}, such an overview would seriously increase the degree of transparency and, thus, would benefit not only the user, but also TikTok.

The impressive accuracy of TikTok's recommender system (RS) mentioned by the literature (e.g. \cite{zhao2021analysisdouyin, AndersonTikTok2020, klug2021trick, chen2019study}), could be used to effectively communicate important messages such as those on COVID-19 countermeasures \cite{basch2020covid}, or place appropriate advertisements. However, such tools can also be easily misused for political manipulation \cite{woolley2016automating}, \cite{TikTokPoliticalAnalysisSerrano2020}, \cite{howard2016bots} or distributing hate speech \cite{weimann2020research}. This can be exacerbated by the closed-loop relationship between users' addiction to the platform and algorithmic optimization \cite{zhao2021analysisdouyin} or filter bubbles. Our hashtag similarity analysis and the analysis of location and language-based differences imply the existence of such filter bubbles both at the level of individual interests but also at a macrolevel related to one's location. The findings of WSJ's investigation \cite{WSJ2021TikTok} also lend evidence to the formation of filter bubbles on TikTok. We therefore propose to countermeasure the creation of filter bubbles not only with recommendation novelty, but also by providing more serendipitous recommendations as this leads to higher perceived preference fit and enjoyment while serving the ultimate goal of increasing the diversity of the recommended content \cite{matt2014escaping}.

\section{Conclusion}
With this work, we aim to contribute to the increase in transparency of how the distribution of content on TikTok is influenced by users' actions or characteristics by identifying the influence of certain factors. We have implemented a sock-puppet auditing technique to interact with the web-version of TikTok mimicking a human user, while collecting data of every post that was encountered. Through this approach we were able to test and analyse the affect of the language and location used to access TikTok, follow- and like-feature, as well as how the recommended content changes as a user watches certain posts longer than others. Our results revealed that all tested factors have an effect on the way TikTok’s RS recommends content to its users. We have also shown that the follow-feature influences the recommendation algorithm the strongest, followed by the video view rate and like feature; besides, we found that the location is a stronger influential factor than the language that is used to access TikTok. Of course, this analysis is not exhaustive and includes only the most explicit factors, while the algorithm without a doubt can be influenced by many other aspects such as, for instance, users' commenting or sharing actions. Nonetheless, with this work we hope to lay the foundation for future research on TikTok's RS that could examine other factors that can influence the algorithm as well as analyze the connection between the RS and the potential for the formation of filter bubbles and the distribution of problematic content on the platform in greater detail.

\section{Acknowledgements}
We thank Prof. Dr. Anikó Hannák for helpful feedback and suggestions on this manuscript. We also thank the Social Computing Group of the University of Zurich for providing the resources necessary to conduct the study. Further, we are grateful to Jan Scholich for his advice on the data analysis implementation.

\bibliographystyle{ACM-Reference-Format}
\bibliography{bibliography} 

\clearpage
\onecolumn
\appendix

\section{Experimental Scenario Details}
\begin{table}[!htb]
    \centering
    \caption{Different experimental groups and their individual scenarios: controlling against noise, language and location, like feature, follow feature, video view rate feature. The yellow highlighted users are the active users and red highlighted scenarios correspond to the failed ones.}
    \begin{tabular}{|p{0.12\textwidth}|p{0.18\textwidth}|p{0.6\textwidth}|}
        \hline
         Test Scenario ID & User IDs & Test Details \\
        \hline\hline
        1 & 72, 73 & Control: collecting 5 batches, collecting\textunderscore data\textunderscore for\textunderscore first\textunderscore posts = True \\
        \hline
        2 & 74, 75 & Control: collecting 5 batches \\
        \hline
        \colorbox{red}{3} & 93, 94 & Control: collecting 5 batches, collecting\textunderscore data\textunderscore for\textunderscore first\textunderscore posts = True \\
        \hline
        4 & 95, 96 & Control: collecting 5 batches \\
        \hline
        5 & 125, 126 & Control : collecting\textunderscore data\textunderscore for\textunderscore first\textunderscore posts = True \\
        \hline
        6 & 137, 138 & Control \\
        \hline
        7 & 139, 140 & Control: collecting\textunderscore data\textunderscore for\textunderscore first\textunderscore posts = True \\
        \hline
        8 & 141, 142 & Control \\
        \hline
        9 & 143, 144 & Control \\
        \hline
        10 & 147, 148 & Control: reuse\textunderscore cookies = True \\
        \hline
        11 & 149, 150 & Control: reuse\textunderscore cookies = True \\
        \hline
        12 & 97, 98, 99, 100 & Language = English; Location = United States and Canada \\
        \hline
        \colorbox{red}{13} & 101, 102, 105, 106 & Language = English; Location = United States and Canada \\
        \hline
        14 & 103, 104, 107, 108 & Language = English and German; Location = United States and Germany \\
        \hline
        15 & 109, 110, 129, 132, 130, 133, 131, 134 & Language = German, English, Spanish, French; Location = United States \\
        \hline
        16 & \colorbox{yellow}{45}, 46 & Randomly liking 6 posts in batch 2, 3, 4, collecting 5 batches \\
        \hline
        17 & \colorbox{yellow}{59}, 60 & Randomly liking 6 posts in batch 2, 3, 4, collecting 5 batches \\
        \hline
        18 & \colorbox{yellow}{61}, 62 & Liking posts based on the user's persona defined by hashtags, collecting 5 batches \\
        \hline
        19 & \colorbox{yellow}{63}, 64 & Liking posts based on the user's persona defined by hashtags, collecting 5 batches \\
        \hline
        20 & \colorbox{yellow}{70}, 71 & Liking posts based on the user's persona defined by hashtags, collecting 5 batches \\
        \hline
        21 & \colorbox{yellow}{123}, 124 & Liking posts based on the user's persona defined by hashtags \\
        \hline
        22 & \colorbox{yellow}{159}, 160 & Liking posts based on the user's persona defined by hashtags, reuse\textunderscore cookies = True \\
        \hline
        23 & \colorbox{yellow}{113}, 114 & Liking posts of specific content creators \\
        \hline
        24 & \colorbox{yellow}{135}, 136 & Liking posts of specific content creators \\
        \hline
        25 & \colorbox{yellow}{115}, 116 & Liking posts with specific sound \\
        \hline
        26 & \colorbox{yellow}{117}, 118 & Liking posts with specific sound \\
        \hline
        27 & \colorbox{yellow}{47}, 48 & Follow a random content creator \\
        \hline
        28 & \colorbox{yellow}{49}, 50 & Follow a random content creator\\
        \hline
        \colorbox{red}{29} & \colorbox{yellow}{51}, 52 & Follow a random content creator \\
        \hline
        30 & \colorbox{yellow}{53}, 54 & Follow a random content creator \\
        \hline
        31 & \colorbox{yellow}{153}, 154 & Follow a random content creator, reuse\textunderscore cookies = True \\
        \hline
        32 & \colorbox{yellow}{155}, 156 & Follow a random content creator, reuse\textunderscore cookies = True \\
        \hline
        33 & \colorbox{yellow}{77}, 78 & VVR: watching 10 random posts for 25\% of their entire length \\
        \hline
        34 & \colorbox{yellow}{79}, 80 & VVR: watching 10 random posts for 50\% of their entire length \\
        \hline
        35 & \colorbox{yellow}{81}, 82 & VVR: watching 10 random posts for 75\% of their entire length \\
        \hline
        36 & \colorbox{yellow}{83}, 84 & VVR: watching 10 random posts for 100\% of their entire length \\
        \hline
        37 & \colorbox{yellow}{85}, 86 & VVR: watching 10 random posts for 200\% of their entire length \\
        \hline
        38 & \colorbox{yellow}{87}, 88 & VVR: watching posts matching user persona for 50\% of their entire length \\
        \hline
        39 & \colorbox{yellow}{145}, 146 & VVR: watching posts matching user persona for 75\% of their entire length \\
        \hline
        40 & \colorbox{yellow}{91}, 92 & VVR: watching posts matching user persona for 100\% of their entire length \\
        \hline
        41 & \colorbox{yellow}{151}, 152 & VVR: watching posts matching user persona for 400\% of their entire length, reusing\textunderscore cookies = true \\
        \hline
        42 & \colorbox{yellow}{157}, 158 & VVR: watching posts matching user persona for 400\% of their entire length, reusing\textunderscore cookies = true, time\textunderscore to\textunderscore look\textunderscore at\textunderscore post\textunderscore normal = 0.5 \\
        \hline
    \end{tabular}
    \label{tab:testgroupdetails}
\end{table}

\section{Difference Analysis Results}
\label{section:differenceanalysisresults}
\begin{table}[!htb]
    \centering
    \caption{Overview of average analysis metrics comparing control and like test scenarios.}
    \begin{tabular}{|l|lll|lll|}
    \hline
    \multirow{2}{*}{Avg. Trend Line Slopes} & \multicolumn{3}{l|}{Control Scenarios}                                   & \multicolumn{3}{l|}{Like Test Scenarios}                                 \\ \cline{2-7} 
                                            & \multicolumn{1}{l|}{3 Batches} & \multicolumn{1}{l|}{5 Batches} & All    & \multicolumn{1}{l|}{3 Batches} & \multicolumn{1}{l|}{5 Batches} & All    \\ \hline
    Diff. Posts                             & \multicolumn{1}{l|}{0.42\%}    & \multicolumn{1}{l|}{1.01\%}    & 0.59\% & \multicolumn{1}{l|}{0.82\%}    & \multicolumn{1}{l|}{0.88\%}    & 0.92\% \\ \hline
    Diff. Hashtags                          & \multicolumn{1}{l|}{0.28\%}    & \multicolumn{1}{l|}{0.98\%}    & 0.65\% & \multicolumn{1}{l|}{0.36\%}    & \multicolumn{1}{l|}{0.77\%}    & 0.65\% \\ \hline
    Diff. Content Creator                   & \multicolumn{1}{l|}{0.23\%}    & \multicolumn{1}{l|}{0.8\%}     & 0.73\% & \multicolumn{1}{l|}{0.72\%}    & \multicolumn{1}{l|}{0.73\%}    & 0.73\% \\ \hline
    Diff. Sounts                            & \multicolumn{1}{l|}{0.4\%}     & \multicolumn{1}{l|}{0.54\%}    & 0.53\% & \multicolumn{1}{l|}{0.78\%}    & \multicolumn{1}{l|}{0.82\%}    & 0.87\% \\ \hline
    \end{tabular}
    \label{tab:likeaveragedresults}
    \vspace{-6mm}
\end{table}

\begin{table}[!htb]
    \centering
    \caption{Overview of average analysis metrics comparing control and follow test scenarios.}
    \begin{tabular}{|l|ll|ll|}
    \hline
    \multirow{2}{*}{Avg. Trend Line Slopes} & \multicolumn{2}{l|}{Control Scenarios}  & \multicolumn{2}{l|}{Follow Test Scenarios} \\ \cline{2-5} 
                                            & \multicolumn{1}{l|}{3 Batches} & All    & \multicolumn{1}{l|}{3 Batches}   & All     \\ \hline
    Diff. Posts                             & \multicolumn{1}{l|}{0.42\%}    & 0.59\% & \multicolumn{1}{l|}{2.03\%}      & 1.59\%  \\ \hline
    Diff. Hashtags                          & \multicolumn{1}{l|}{0.28\%}    & 0.65\% & \multicolumn{1}{l|}{1.79\%}      & 1.46\%  \\ \hline
    Diff. Content Creator                   & \multicolumn{1}{l|}{0.23\%}    & 0.42\% & \multicolumn{1}{l|}{1.73\%}      & 1.3\%   \\ \hline
    Diff. Sounds                            & \multicolumn{1}{l|}{0.4\%}     & 0.53\% & \multicolumn{1}{l|}{1.89\%}      & 1.53\%  \\ \hline
    \end{tabular}
    \label{tab:followaveragedresults}
    \vspace{-6mm}
\end{table}

\begin{table}[!htb]
    \centering
    \caption{Overview of average analysis metrics comparing control and VVR test scenarios.}
    \begin{tabular}{|l|ll|llll|}
    \hline
    \multirow{2}{*}{Avg. Trend Line Slopes} & \multicolumn{2}{l|}{Control Scenarios}  & \multicolumn{4}{l|}{VVR Test Scenarios}                                                              \\ \cline{2-7} 
                                            & \multicolumn{1}{l|}{3 Batches} & All    & \multicolumn{1}{l|}{3 Batches} & \multicolumn{1}{l|}{All}    & \multicolumn{1}{l|}{Random} & Persona \\ \hline
    Diff. Posts                             & \multicolumn{1}{l|}{0.42\%}    & 0.59\% & \multicolumn{1}{l|}{0.75\%}    & \multicolumn{1}{l|}{0.98\%} & \multicolumn{1}{l|}{0.67\%} & 0.95\%  \\ \hline
    Diff. Hashtags                          & \multicolumn{1}{l|}{0.28\%}    & 0.65\% & \multicolumn{1}{l|}{0.62\%}    & \multicolumn{1}{l|}{0.82\%} & \multicolumn{1}{l|}{0.59\%} & 0.69\%  \\ \hline
    Diff. Content Creator                   & \multicolumn{1}{l|}{0.23\%}    & 0.42\% & \multicolumn{1}{l|}{0.51\%}    & \multicolumn{1}{l|}{0.63\%} & \multicolumn{1}{l|}{0.41\%} & 0.75\%  \\ \hline
    Diff. Sounds                            & \multicolumn{1}{l|}{0.4\%}     & 0.53\% & \multicolumn{1}{l|}{0.64\%}    & \multicolumn{1}{l|}{0.84\%} & \multicolumn{1}{l|}{0.58\%} & 0.81\%  \\ \hline
    \end{tabular}
    \label{tab:vvraveragedresults}
\end{table}

\section{Additional Figures}
\begin{figure*}[h]
    \centering
    \begin{minipage}{0.45\textwidth}
        \includegraphics[width=\textwidth]{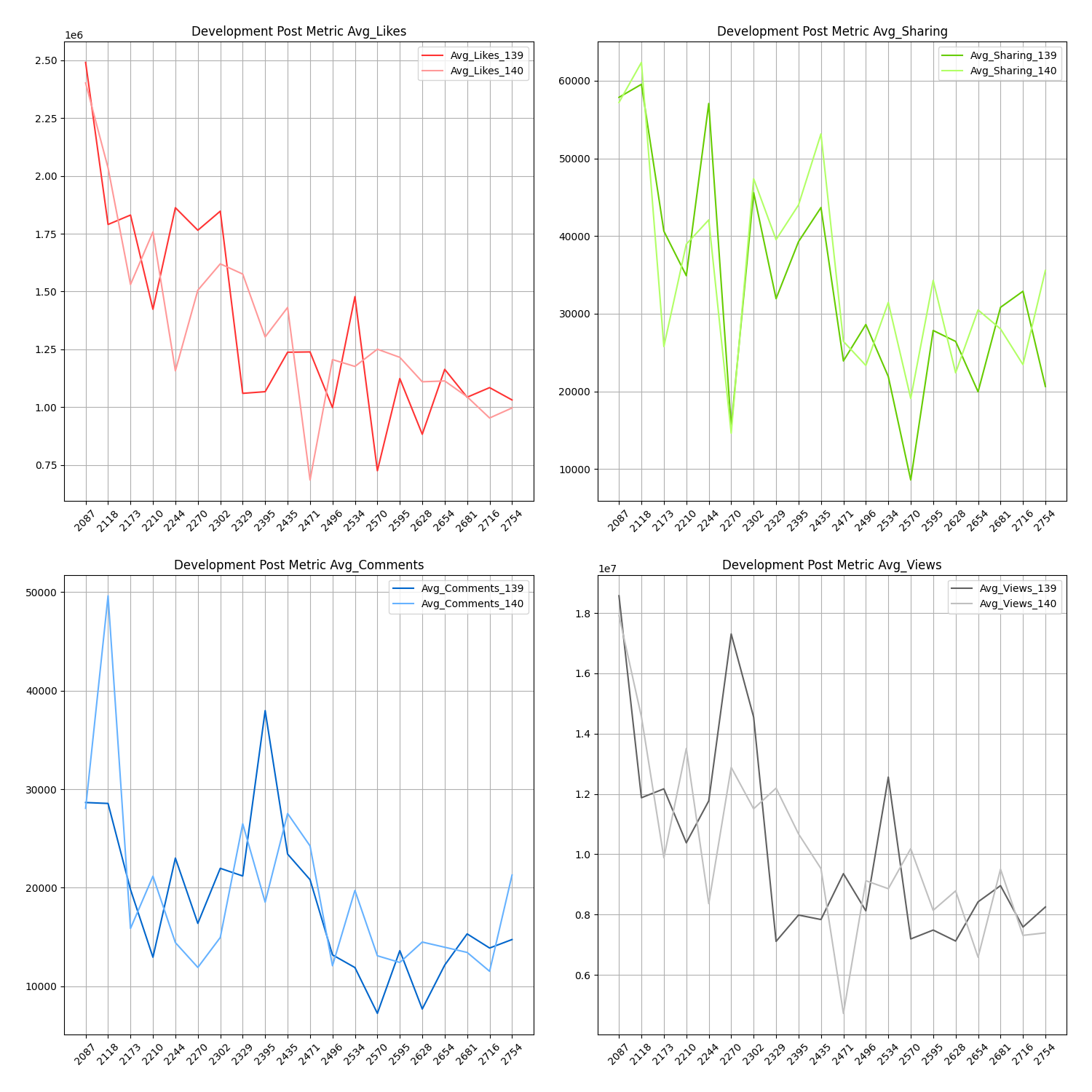}
        \caption{Post metrics (Likes-Shares-Comments-Views) changes for test scenario 7.}
        \label{fig:postmetricscontrolscenario7}
    \end{minipage}
    \hspace{2em}
    \begin{minipage}{0.45\textwidth}
        \includegraphics[width=\textwidth]{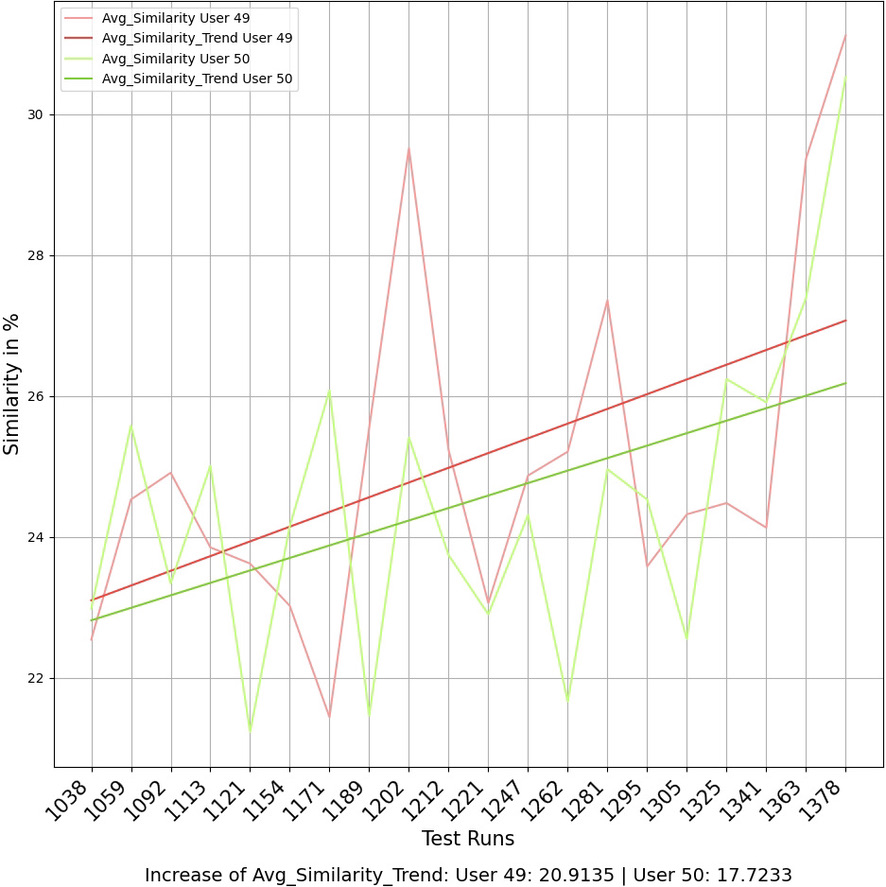}
        \caption{Hashtag similarity within feed of each user per test run for scenario 28.}
        \label{fig:followresultsdiffhashtagsim}
    \end{minipage}
\end{figure*}

\end{document}